\documentclass[draft]{agujournal2019}
\usepackage{url} 
\usepackage{lineno}
\usepackage[inline]{trackchanges} 
\usepackage{soul}
\usepackage{amsmath,amssymb}
\usepackage[utf8]{inputenc}
\draftfalse

\journalname{Geophysical Research Letters}

\begin{document}

%
%


\title{Learning Climate Variability from Scarce Data with Diffusion Models: An Idealised Test Case for ENSO}

%
%




\authors{Lluís Palma\affil{1,2}, Vincent Verjans\affil{1}, Amanda Duarte\affil{1}, Albert Soret\affil{1} and Markus Donat\affil{1,3}}


\affiliation{1}{Barcelona Supercomputing Center, Earth Sciences Department, Barcelona, Spain}
\affiliation{2}{Facultat de Física, Universitat de Barcelona, Barcelona, Spain}
\affiliation{3}{Institució Catalana de Recerca i Estudis Avançats (ICREA), Barcelona, Spain}





\correspondingauthor{Lluís Palma}{lluis.palma@bsc.es}



\begin{keypoints}
\item We use synthetic tropical Pacific SST data to test how accurately diffusion models learn the distribution of the data they are trained on
\item Diffusion models succeed given enough data but fail to converge to optimal results for records of observational length
\item Regularisation avoids overfitting yet neither it nor pre-training overcomes the limited sampling of variability in the observational record
\end{keypoints}

%
%

%
%


\begin{abstract}

We test how faithfully diffusion models can map the distribution of tropical Pacific sea surface temperature (SST) monthly anomalies. Using long simulations from Gaussian and non-Gaussian Linear Inverse Models (LIMs) as an idealised ground truth, we perform data ablation experiments. Given sufficient training data, a diffusion model recovers their central statistics --- the spatial patterns of variability, the partition of variance among them, and the El Ni\~no/La Ni\~na asymmetry --- although only specific diffusion parameterisations do so robustly. For records as short as the ERSSTv5 observational record, however, the model falls short of these results. Regularisation prevents overfitting, yet neither regularisation nor pre-training on CMIP6 simulations compensates for the limited sampling of variability in the observational record, which affects higher-order moments most. Synthetic data with known statistics thus offer a controlled benchmark for exposing the limitations of generative models and related machine learning methods in climate applications.

\end{abstract}

\section*{Plain Language Summary}

  El Ni\~no and La Ni\~na are major climate phenomena that affect weather patterns worldwide, yet we have only about 70 years of detailed sea surface temperature observations --- too few for modern AI methods, which typically need thousands of examples to learn reliably. We tested whether diffusion models, a type of AI originally developed for image generation, can learn the essential patterns of tropical Pacific variability, using very long synthetic datasets whose characteristics are known. These models succeed when given enough data, but with a record as short as the observations they miss finer features, such as how often and how strongly extreme events occur. Training the model first on many climate simulations and then adapting it to observations helps when data are extremely scarce, but cannot make up for the variability that a short record does not contain. Synthetic data thus offer a way to test what such AI methods can and cannot learn before they are applied to real climate data.

%
%

%


%
%
%
%

\section{Introduction}

Deep learning has rivalled traditional methods in many weather and climate applications, from prediction \cite{pathakFourCastNetGlobalDatadriven2022,lamLearningSkillfulMediumrange2023,kochkovNeuralGeneralCirculation2024} to emulation \cite{watt-meyerACE2AccuratelyLearning2025} and downscaling \cite{aichConditionalDiffusionModels2026,springenbergDiffScaleContinuousDownscaling2026}. For most climate applications, however, progress is hampered by the short span of reanalyses available for training (40--70 years): unlike weather forecasting, where high-frequency variability yields many independent samples over short periods, slower variability yields far fewer samples \cite{gibsonTrainingMachineLearning2021,miloshevichProbabilisticForecastsExtreme2023,materiaArtificialIntelligenceClimate2024}. One example is the El Ni\~no--Southern Oscillation (ENSO), the dominant mode of interannual variability in the tropical Pacific \cite{wangNinoSouthernOscillation2017}. ENSO recurs every 2--7 years, alternating between anomalous warming (El Ni\~no) and cooling (La Ni\~na) of the central and eastern equatorial Pacific. ENSO-driven anomalies alter tropical convection and trigger atmospheric teleconnections that drive temperature and precipitation impacts in large parts of the world, with major societal consequences \cite{mcphadenENSOIntegratingConcept2006,liuNonlinearNinoImpacts2023}. Yet a single reanalysis may contain only a few dozen ENSO cycles, limiting the use of deep learning models with many trainable parameters.

Some deep learning approaches address this scarcity by training on climate model output \cite{anderssonSeasonalArcticSea2021,gibsonTrainingMachineLearning2021,panImprovingSeasonalForecast2022,palmaDatadrivenSeasonalClimate2026}, benefiting from thousands of simulated years that span regime shifts and unobserved trends, though at the cost of inheriting model biases, such as misrepresented teleconnections and unresolved small-scale processes. An emerging alternative trains AI models on higher-frequency reanalysis data and runs them autoregressively to reach longer timescales \cite{watt-meyerACE2AccuratelyLearning2025,kentSkilfulGlobalSeasonal2025}, avoiding model biases but not the limited sampling of low-frequency variability in the training data. Across all approaches, complex architectures make it difficult to verify that learned relationships are physically consistent, with skill scores and summary statistics typically serving as proxy validation \cite{raspWeatherBench2Benchmark2024}.

Among deep learning methods, diffusion models \cite{hoDenoisingDiffusionProbabilistic2020,songScoreBasedGenerativeModeling2021,karrasElucidatingDesignSpace2022,lipmanFlowMatchingGenerative2023} stand out for their ability to model the nonlinear, non-Gaussian behaviour intrinsic to the climate system. As generative models, they learn the distribution of their training data, whose true form is unknown and of which only a finite sample is observed. Their progressive denoising objective, which requires producing coherent fields from heavily corrupted to nearly clean inputs, helps prevent overfitting and avoids the oversmoothed fields typical of deterministic models trained to minimise RMSE. Applications then condition this distribution on an input: on the initial state for ensemble weather forecasts \cite{priceGenCastDiffusionbasedEnsemble2024}, on large-scale information such as sea surface temperatures for climate emulation \cite{bassettiDiffESMConditionalEmulation2024,manshausenAccurateExtremeEvent2026}, and on coarse-resolution fields for downscaling and bias adjustment \cite{aichConditionalDiffusionModels2026,wanRegionalClimateRisk2026}. Marginalising these conditional distributions over all inputs yields the unconditional one, so all these applications rely on the latter being modelled faithfully. However, how faithfully diffusion models capture it, and how much data they need to do so, is hard to verify with short reanalysis records, in which trends and overlapping modes also confound the signal.

We therefore leverage a synthetic multi-millennial SST dataset from a low-order ENSO model \cite{martinez-villalobosLowOrderDataDrivenModel2025}, a simple statistical process whose generating distribution is known exactly. This lets us test whether a diffusion model recovers the statistics central to that process, namely the spatial patterns of temporal variability, the partition of variance among them, and the non-Gaussian features, in a controlled setting free from trends, distributional shifts and higher-order modes. We focus on the unconditional distribution, leaving temporal dynamics and initialised predictions for future work.

\section{Data and Methods}\label{sec:data}

\subsection{Data}


To learn ENSO variability, we use 120k monthly tropical Pacific (20\textdegree S--20\textdegree N, 120\textdegree E--50\textdegree W) SST fields generated from two sets of Linear Inverse Models (LIMs; \citeA{penlandRandomForcingForecasting1989,martinez-villalobosLowOrderDataDrivenModel2025,martinez-villalobosTropicalPacificSST2026}). Specifically, we use both the standard LIM and non-Gaussian (NG-LIM) versions \cite{martinez-villalobosLowOrderDataDrivenModel2025}. In a standard stationary LIM, the evolution of the tropical Pacific state vector is simulated as follows:

\begin{equation}
\frac{d\mathbf{x}}{dt} = \mathbf{M}\mathbf{x} + \mathbf{B}\boldsymbol{\eta}, \label{eq:lim}
\end{equation}
where $\mathbf{x}$ is a 10-dimensional state vector whose first two components are the East and Central Pacific (EP, CP) indices, defined as linear combinations of the leading two Principal Components (PCs) of tropical Pacific SST variability, and whose remaining eight components are PCs 3--10. The matrix $\mathbf{M}$ represents the linearized deterministic dynamics derived from the empirical 1 month lag covariance matrix \cite{penlandOptimalGrowthTropical1995}, $\mathbf{B}$ is the noise covariance matrix derived from the fluctuation-dissipation relationship \cite{penlandBalanceConditionStochastic1994}, and $\boldsymbol{\eta}$ is a Gaussian white noise vector. In the NG-LIM, a Yeo-Johnson transformation \cite{yeoNewFamilyPower2000} is applied as a pre-processing step that maps the skewed state variables $\mathbf{x}$ to near-Gaussianity; the linear dynamics of Equation~\ref{eq:lim} are then calibrated and integrated in this transformed space, and the inverse transformation maps the integrated state back to the original (untransformed) variables, thereby preserving the ENSO nonlinearities and asymmetries not captured by the standard LIM \cite{choiENSOTransitionDuration2013, dinezioNonlinearControlsPersistence2014,martinez-villalobosObservedNinoLaNina2019}. The gridded SST anomaly fields are obtained by mapping the 10-dimensional state onto the latitude--longitude grid through the corresponding 10 EOFs, which together retain 90\% of the ERSSTv5 tropical Pacific SST variance. Thus, although each synthetic field has $D = 16 \times 76 = 1216$ grid points (1009 in experiments involving ERSSTv5 and CMIP6, which share a common ocean mask), both the LIM and NG-LIM datasets lie by construction on a known subspace of dimension $d = 10$.

We use the LIM and NG-LIM realisations published by \citeA{martinez-villalobosLowOrderDataDrivenModel2025}, both calibrated to the 1948--2022 ERSSTv5 observations \cite{huangExtendedReconstructedSea2017}. We also use the raw ERSSTv5 SST fields and NG-LIM emulations of 30 CMIP6 \cite{eyringOverviewCoupledModel2016} models from \citeA{martinez-villalobosTropicalPacificSST2026}, calibrated over the same period.

\subsection{Diffusion Model}

Diffusion models (DMs; \citeA{hoDenoisingDiffusionProbabilistic2020,songScoreBasedGenerativeModeling2021}), inspired by non-equilibrium thermodynamics \cite{sohl-dicksteinDeepUnsupervisedLearning2015}, follow a two-stage process: a forward process that gradually corrupts data with Gaussian noise, and a reverse process that generates new samples by denoising. More formally, the DM forward process follows:

\begin{equation}
\mathbf{z}_t = \alpha_t \mathbf{x} + \sigma_t \boldsymbol{\epsilon}, \label{eq:forward}
\end{equation}
where $\alpha_t$ is the signal scaling coefficient at time $t$, controlling how much of the original data $\mathbf{x}$ is preserved, and $\sigma_t$ controls the noise level ($\boldsymbol{\epsilon} \sim \mathcal{N}(0, \mathbf{I})$). Note that diffusion time $t$ indexes the noising process, not the physical time over which weather evolves, from $t=1$ for the original data to $t=0$ for pure noise. As $t$ decreases, the ratio $\alpha_t / \sigma_t$ decreases monotonically, until the data are indistinguishable from noise. These marginals form a time-continuous, increasingly noisy sequence — the probability path:

\begin{equation}
p(\mathbf{z}_t) = \int p(\mathbf{z}_t \mid \mathbf{x})\, p(\mathbf{x})\, d\mathbf{x} = \int \mathcal{N}(\mathbf{z}_t\mid\alpha_t \mathbf{x}, \sigma_t^2 \mathbf{I})\, p(\mathbf{x})\, d\mathbf{x}. \label{eq:probpath}
\end{equation}

We adopt the ordinary differential equation (ODE)-based \cite{songScoreBasedGenerativeModeling2021} formulation of flow matching \cite{lipmanFlowMatchingGenerative2023}; \citeA{laiPrinciplesDiffusionModels2025} give an overview of related diffusion formalisms. We parameterise the time evolution of $p(\mathbf{z}_t)$ via the time derivative of $\mathbf{z}$, a velocity field $\mathbf{v}$, which prescribes how samples must evolve so that their distribution matches $p(\mathbf{z}_t)$ at every $t$ --- a constraint formalised by the continuity equation. For the Gaussian path defined by Eq.~\ref{eq:probpath}, the corresponding probability flow ODE can be expressed as:

\begin{equation}
\frac{d\mathbf{z}_t}{dt} = \mathbf{v}_t(\mathbf{z}_t). \label{eq:velocity}
\end{equation}

We adopt a simple linear schedule \cite{lipmanFlowMatchingGenerative2023}, for which $\alpha_t = t$ and $\sigma_t = 1 - t$. The objective is then to fit the trainable parameters $\theta$ of a neural network $\mathbf{v}_\theta(\mathbf{z}_t, t) \approx \mathbf{v}_t(\mathbf{z}_t)$ that approximates this velocity field. As $\mathbf{v}_t(\mathbf{z}_t)$ is generally intractable, we seek to minimise the network output against the conditional velocity $\mathbf{v}_t(\mathbf{z}_t \mid \mathbf{x})$, which can be derived from the time derivative of Equation~\ref{eq:forward} as:

\begin{equation}
\mathbf{v}_t(\mathbf{z}_t \mid \mathbf{x}) = \alpha'_t \mathbf{x} + \sigma'_t \boldsymbol{\epsilon} = \mathbf{x} - \boldsymbol{\epsilon} = \frac{\mathbf{x} - \mathbf{z}_t}{1-t} = \frac{\mathbf{z}_t - \boldsymbol{\epsilon}}{t},
  \label{eq:cond_vel}
\end{equation}
yielding the conditional (on $\mathbf{x}$) flow matching loss:

\begin{equation}
\mathcal{L} = \mathbb{E}_{t \sim \mathcal{U}(0,1),\, \mathbf{x} \sim p(\mathbf{x}),\, \boldsymbol{\epsilon} \sim \mathcal{N}(0,\mathbf{I})} \left[\left\| \mathbf{v}_\theta(\mathbf{z}_t, t) - \mathbf{v}_t(\mathbf{z}_t \mid \mathbf{x}) \right\|^2 \right]. \label{eq:loss}
\end{equation}

\citeA{lipmanFlowMatchingGenerative2023} shows that this loss shares the same gradient with respect to $\theta$ as the marginal flow matching loss (Mean Squared Error against $\mathbf{v}_t(\mathbf{z}_t)$), so minimising one also minimises the other. Once $\mathbf{v}_\theta$ is learned, the generative process recovers samples from the data distribution: we integrate Equation~\ref{eq:velocity} from $t=0$ to $t=1$, replacing $\mathbf{v}_t(\mathbf{z}_t)$ with $\mathbf{v}_\theta(\mathbf{z}_t, t)$.

Given Equation~\ref{eq:cond_vel}, the network, whose raw output we denote $\mathbf{d}_\theta$, can be trained to predict any of three equivalent targets at each time step $t$: the clean data, $\mathbf{d}_\theta \approx \mathbf{x}$, from which the velocity is recovered as $\mathbf{v}_t = (\mathbf{d}_\theta - \mathbf{z}_t)/(1-t)$, close to the original diffusion formulation \cite{sohl-dicksteinDeepUnsupervisedLearning2015}; the noise, $\mathbf{d}_\theta \approx \boldsymbol{\epsilon}$, with $\mathbf{v}_t = (\mathbf{z}_t - \mathbf{d}_\theta)/t$, as in Denoising Diffusion Probabilistic Models (DDPM) and the closely related score-based models \cite{hoDenoisingDiffusionProbabilistic2020,songGenerativeModelingEstimating2020}; or the velocity itself, $\mathbf{d}_\theta \approx \mathbf{v}_t$ \cite{salimansProgressiveDistillationFast2022, lipmanFlowMatchingGenerative2023}. These are related by simple identities and differ in training only by time-dependent reweighting factors \cite{laiPrinciplesDiffusionModels2025,liBackBasicsLet2025}.

In practice, however, these parameterisations behave very differently. \citeA{liBackBasicsLet2025} argue that natural data occupy a low-dimensional manifold within the high-dimensional observation space, while noise spans the full space. A network predicting clean data ($\mathbf{x}$-prediction) therefore only needs to capture this low-dimensional structure, whereas predicting noise ($\boldsymbol{\epsilon}$-prediction), or an intermediate quantity ($\mathbf{v}$-prediction), requires preserving high-dimensional information. We hypothesise that this reasoning extends to climate fields, where a low-dimensional structure can explain a large fraction of climate variability \cite{devironGlobalModesClimate2013}. 

Thus, our model ingests monthly SST anomaly fields from the tropical Pacific ($\mathbf{x} \in \mathbb{R}^{n_{lat} \times n_{lon}}$) and learns the velocity field ($\mathbf{v}_\theta \in \mathbb{R}^{n_{lat} \times n_{lon}}$) that transports Gaussian noise to the distribution of SST anomalies in the region (architecture, training and sampling in Text~S1). All models are trained without dropout or weight decay, using early stopping on the last 20\% of each training record, which acts as the main implicit regulariser of diffusion models \cite{farghlyBenignOverfittingDoes2026}.

\section{Results}\label{sec:results}

To evaluate whether diffusion models learn the generating distribution, we start by training our model ($\mathbf{x}$-prediction) separately on the LIM and NG-LIM synthetic datasets, obtained by stochastically integrating the ERSSTv5-calibrated models (Section~\ref{sec:data}). Each synthetic dataset is split in time: training months are drawn from the first 96,000, and all metrics are computed on the last 24,000, which are used neither for training nor for early stopping (Text~S1.3). Here and throughout, ``reference'' refers to these held-out months, against which the diffusion model is evaluated, whereas ``generated'' refers to samples drawn from the trained diffusion model.

In line with \citeA{martinez-villalobosLowOrderDataDrivenModel2025}, Figure 1 shows the joint PC1–PC2 scatter plots (a, f) and spatial skewness maps for reference (b, g) and generated (c, h) data, testing whether the model preserves the non-Gaussian warm–cold asymmetry characteristic of ENSO. Overall, the diffusion model reproduces both the Gaussian (LIM) and non-Gaussian (NG-LIM) distributions. As the standard LIM has a Gaussian stationary distribution, its true skewness and PC1--PC2 curvature are zero; the faint skewness patterns in (b, c), their low pattern correlation and the slight curvature of the fits in (a) reflect finite sampling (pattern correlation 0.18 between the two LIM skewness maps, against 0.96 for the NG-LIM). When trained on the NG-LIM dataset, generated samples reproduce both the characteristic U shape of the PC1–PC2 distribution (Figure~1f), reflecting the tendency for EP El Niños (positive PC1, positive PC2) and CP La Niñas (negative PC1, positive PC2) to reach larger anomalies \cite{choiENSOTransitionDuration2013,martinez-villalobosObservedNinoLaNina2019}, and the horseshoe-shaped skewness pattern (Figure~1g, h), positive in the eastern and negative in the western Pacific. The EOF cross-correlation matrices (d, i) show that the model recovers the first 10 EOFs of both datasets, with a variance-weighted mean diagonal correlation of 0.97 (LIM) and 0.96 (NG-LIM); the few off-diagonal entries above 0.3 involve neighbouring EOFs of similar variance, which are subject to mode mixing. Projecting the generated data onto the reference EOFs (e, j), the variance carried by each EOF matches the reference partition within $\sim$3 percentage points.

  \begin{figure}
  \noindent\includegraphics[width=\textwidth]{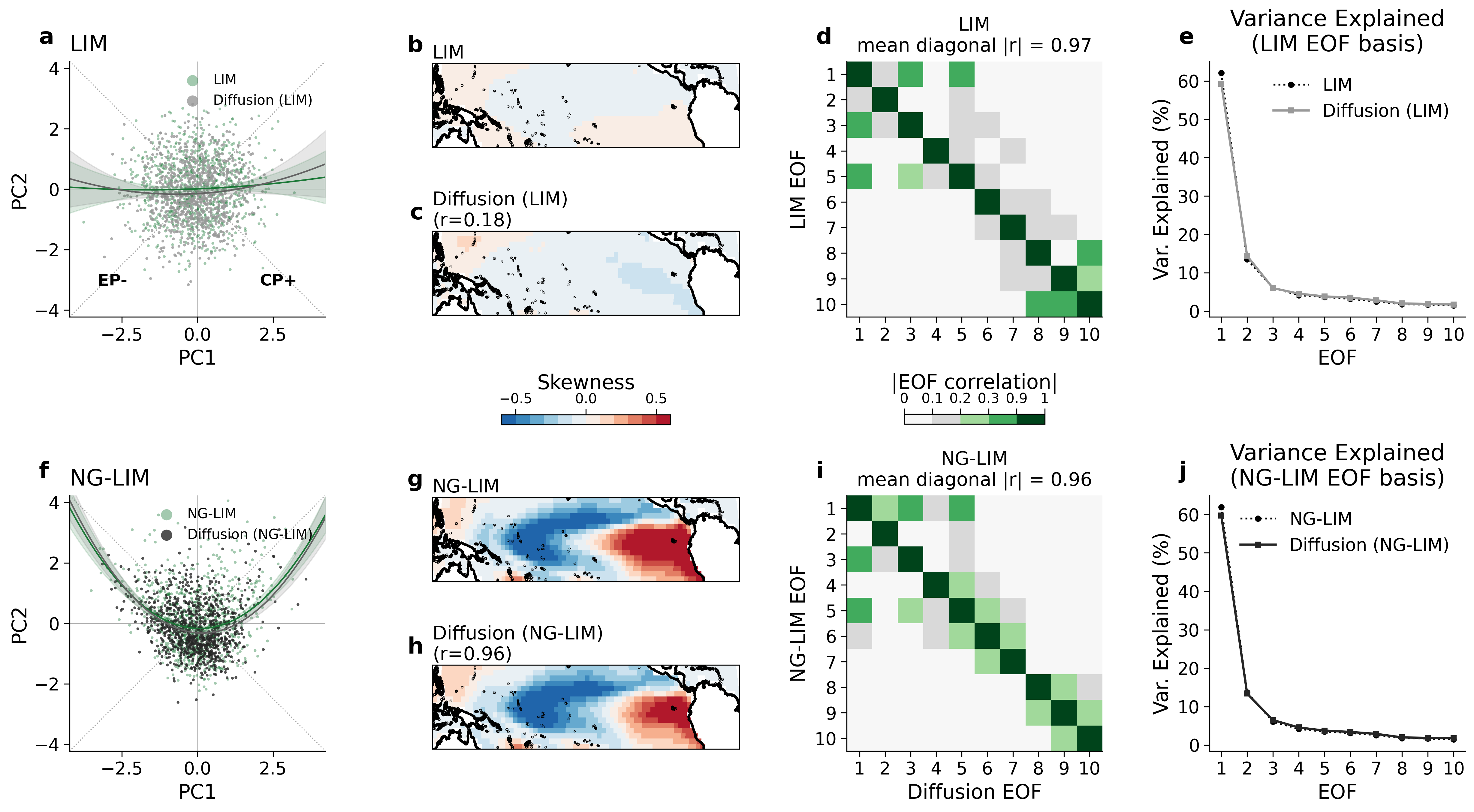}
  \caption{Top row (a--e): diffusion model trained on Linear Inverse Model (LIM) data. Bottom row (f--j): trained on non-Gaussian LIM (NG-LIM) data. In all panels, ``LIM'' and ``NG-LIM'' denote the reference (synthetic) data and ``Diffusion (LIM)'' and ``Diffusion (NG-LIM)'' the samples generated by the corresponding diffusion model. (a,f) Scatter plots of the two leading principal components (PC1 vs PC2) for reference data (green) and generated samples (grey/black), with quadratic fit curves highlighting the nonlinear PC1--PC2 relationship. Scatter shows a random subset of 900 samples per distribution; shading gives the 95\% confidence band of the fit (Text S1.4). (b,c,g,h) Spatial skewness maps for reference data (b,g) and generated samples (c,h). For the LIM (top row), the true skewness is zero, so the residual patterns in (b,c) are sampling noise. (d,i) Cross-correlation between the EOFs of the reference data (rows) and those of the generated samples (columns), computed independently on each dataset (note the nonlinear colour scaling). (e,j) Percentage of variance explained per EOF, with generated samples projected onto the reference EOFs: reference data (dotted) and generated samples (solid).}
  \label{fig:lim_vs_nglim}
  \end{figure}

Motivated by \citeA{liBackBasicsLet2025}, we test on the NG-LIM dataset whether results differ across diffusion parameterisations: $\mathbf{x}$-prediction, $\mathbf{v}$-prediction, and $\boldsymbol{\epsilon}$-prediction \cite{laiPrinciplesDiffusionModels2025}. Figure~S1 shows PC1–PC2 scatter plots, EOF spatial cross-correlation matrices, and projected variance per EOF for each diffusion variant. $\boldsymbol{\epsilon}$-prediction fails all three diagnostics. $\mathbf{v}$-prediction reproduces the PC1–PC2 distribution and EOF correlations well (diagonal mean of 0.95), but projects substantially less variance onto the reference EOF patterns ($\sim$10\% on EOF 1 against $\sim$60\% in the NG-LIM; Figure~S1h), indicating that most of its variance lies outside the 10-dimensional subspace of the data. Thus, only $\mathbf{x}$-prediction fully reproduces the modes of variability of the reference data. 

We further test the sensitivity to the input dimension $D$, the number of values the network ingests and outputs per sample (Figure~S2), reducing it with PCA from the full grid ($D=1216$) down to the manifold dimension ($d=10$). Since the fields have rank 10, PCA with any $D \geq 10$ retains all the variance: the data are identical for every $D$, and the model must recover the same 10-dimensional structure from within a $D$-dimensional space. As $D$ approaches the hidden-layer width (256), $\boldsymbol{\epsilon}$-prediction degrades sharply (already at $D = 200$), and once $D$ exceeds it, $\mathbf{v}$-prediction degrades in projected variance and skewness, whereas $\mathbf{x}$-prediction holds across dimensionalities and diagnostics, consistent with \citeA{liBackBasicsLet2025}. This is an idealised limit, as the synthetic fields have exactly zero variance outside the 10-dimensional subspace; observed SST fields are only approximately low-dimensional, but since most of their variance is concentrated in a few modes, we hypothesise that a similar, if less sharp, behaviour holds for real data.

These results indicate that diffusion models can learn the generating distribution of both Gaussian and non-Gaussian ENSO synthetic datasets when trained on the full 96,000-month training record. We now ask what they can learn from records as short as the observational record ($\sim$720 monthly samples). Since a real record is one contiguous block, we train on consecutive blocks of $N$ months, from 60 up to the full training record, and repeat training 10 times per $N$ with blocks starting at random positions.

Figure 2a--d shows five metrics as a function of $N$ (Text S1.4): the variance-weighted diagonal EOF correlation, the spatial correlations of the skewness and kurtosis patterns, a nearest-neighbour memorisation ratio and a diversity ratio (values near 1 indicate generalisation; Figure~S9 shows the projected variance ratio). To separate what the model fails to learn from what the record itself does not contain, we compute a data ceiling: the same metrics applied to the months the model is fitted on, scored against the same held-out reference (Text~S1.4). Three regimes emerge. In a data-scarce regime (up to a few hundred months), scores are low and the model partly memorises its training block: at $N = 60$ the memorisation ratio drops to 0.48 while the EOF correlation is still 0.71, so standard statistics can look acceptable even for a model that is copying its training data. In a sampling-limited regime, which includes the observational record, the model reaches the data ceiling: at $N = 720$ its median skewness correlation is 0.83 (ceiling 0.84; Table~S2) and its EOF correlation 0.88 (ceiling 0.85). Its spread across blocks, however, is as wide as that of the ceiling itself (skewness correlations 0.65--0.94 for the model against 0.67--0.90 for the ceiling; kurtosis correlations from $-0.18$ to 0.90), so scores on a single record mostly reflect that record. Only in a converged regime, beyond a few thousand months, does this spread collapse (to $\sim$0.1 in skewness correlation and $\sim$0.15 in the variance ratio by $\sim$5,000 months). The sample sizes delimiting these regimes depend on the architecture, the training recipe and the complexity of the target, and are specific to this particular setup. Finally, removing extreme values (by Ni\~no 3.4 index) from the training set (Figure~S3) shows that the model fails to reproduce them: it interpolates within the support of its training data but does not extrapolate beyond it.

  \begin{figure}
  \noindent\includegraphics[width=\textwidth]{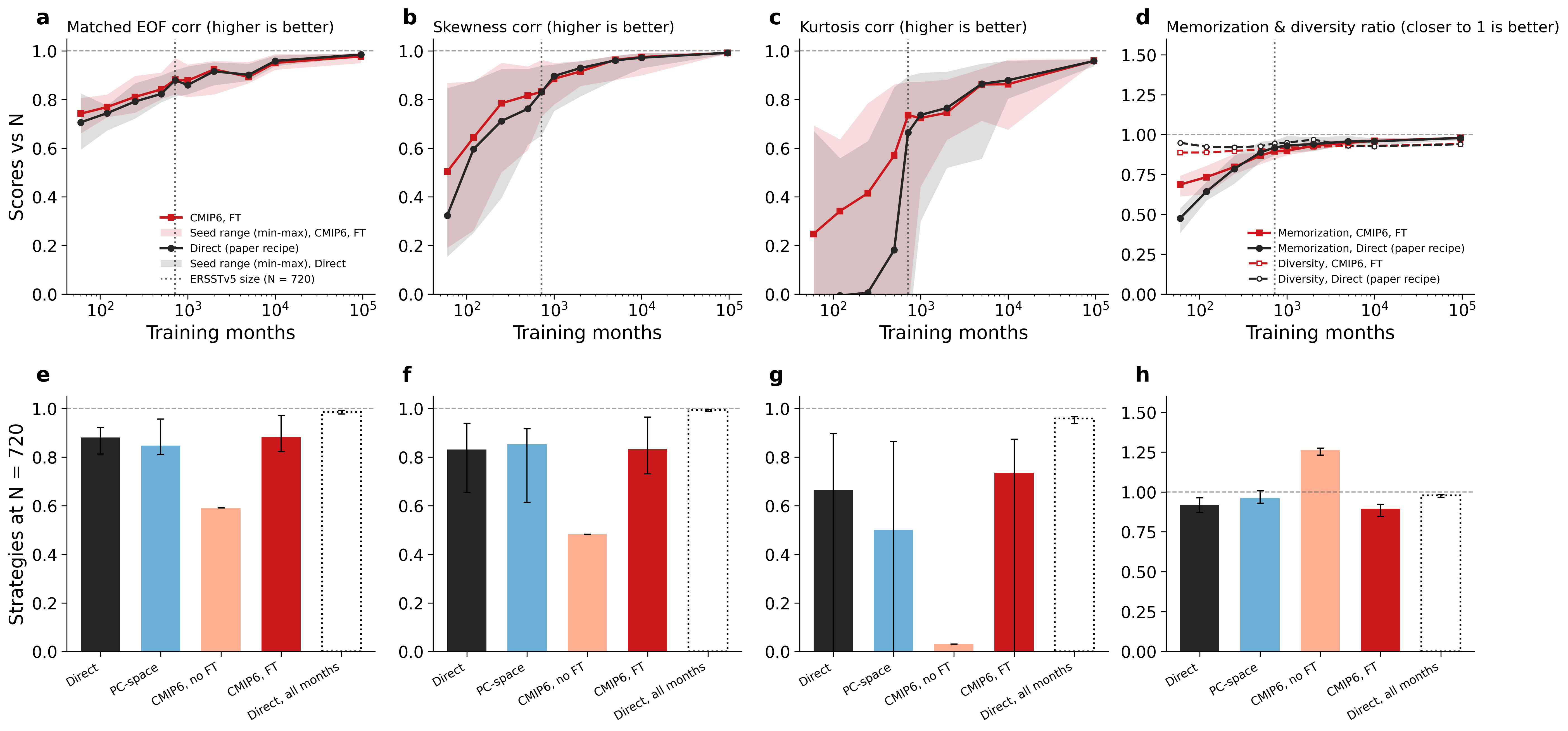}
  \caption{Performance of $\mathbf{x}$-prediction diffusion on the NG-LIM dataset: (a, e) variance-weighted EOF pattern correlation, (b, f) spatial skewness correlation, (c, g) spatial kurtosis correlation and (d, h) memorisation ratio (solid) and diversity ratio (dashed; Text S1.4). Top row (a--d): as a function of the number $N$ of consecutive training months, for direct training (black) and CMIP6 pre-training followed by fine-tuning (red); lines show the median and shading the range (min--max) across 10 seeds, and the vertical dotted line marks $N = 720$, the length of the ERSSTv5 training record. Bottom row (e--h): strategies at $N = 720$: direct training (black), training in a 100-dimensional PC space (light blue), CMIP6 pre-training without fine-tuning (light red) and with fine-tuning (red), and direct training on the full 96,000-month record (white, dotted). Bars show the median and error bars the range across 10 seeds.}
  \label{fig:ablation_vs_N}
  \end{figure}

Several strategies have been proposed to address data scarcity in deep learning, from dimensionality reduction to transfer learning from climate model simulations \cite{anderssonSeasonalArcticSea2021,gibsonTrainingMachineLearning2021,panImprovingSeasonalForecast2022,materiaArtificialIntelligenceClimate2024}. We test whether they help in the sampling-limited regime ($N = 720$). Figure 2e--h compares direct training (black), training in a 100-dimensional PC space (light blue), a model pre-trained on the pooled NG-LIM emulations of 30 CMIP6 models without fine-tuning (light red) and the same model fine-tuned on the 720 target months (red), with direct training on the entire NG-LIM training record ($N = 96{,}000$) as reference (white). The CMIP6 emulations are independent of ERSSTv5, which enters only through the fine-tuning samples, if at all.

Training in PC space neither helps nor hurts (skewness correlation 0.85 vs 0.83). Without fine-tuning, the CMIP6-pre-trained model learns an erroneous variability structure (EOF correlation 0.59, skewness 0.48), reflecting known biases in CMIP6 simulations (Figures~S4--S5); its memorisation ratio above 1 (1.26) reflects the fundamental differences between the CMIP6 emulations and the target distribution. After fine-tuning, it matches direct training at $N = 720$ on every metric (skewness 0.83, EOF 0.88) and barely narrows the spread across blocks. Its benefit is confined to the data-scarce regime (Figure~2a--d), where it raises skewness correlations (0.50 vs 0.32 at $N = 60$) and, above all, kurtosis correlations (0.25--0.57 vs $-0.22$ to 0.18 for direct training, from 60 to 500 months). Explicit regularisation with dropout and weight decay degrades every configuration (e.g. skewness 0.59 vs 0.83 for direct training; Table~S2), consistent with early stopping being the effective regulariser of diffusion models \cite{farghlyBenignOverfittingDoes2026}.

To test whether the fine-tuned model inherits the biases of its pre-training data, we also pre-train on two pools of six CMIP6 emulators whose Ni\~no-3 skewness has the same or the opposite sign as the target's, and evaluate them on two emulators excluded from both pools, MIROC6 and ACCESS-ESM1-5 (a perfect-model setting; Text~S1.5). The pool determines the skewness pattern only before fine-tuning: for MIROC6, the two pools give skewness correlations of 0.87 vs 0.17. After fine-tuning, both converge to the data ceiling (0.82 and 0.79, ceiling 0.79; Figures~S6--S7, Table~S3). In this perfect-model setting the gain from pre-training on short records persists (median paired gain +0.06 to +0.24 at 60--120 months; Figure~S8) and shrinks to $\sim$0.05 at $N = 720$.

Finally, we train on the ERSSTv5 observations themselves rather than on LIM emulations, using the first 720 months (1948--2007) either directly or by fine-tuning the CMIP6 pre-trained model, and hold out 2008--2022 (180 months). Figure 3 compares PC1--PC2 scatter plots (a--d), skewness maps (e--h) and kurtosis maps (i--l) for the NG-LIM, the two diffusion models and the observations. Because 180 months sample these maps poorly (Figure~S10), Figure 3 scores all models against the full 1948--2022 record. These scores are in-sample: the diffusion models were fitted on 720 of its 900 months, and the NG-LIM was calibrated on all of them. Both diffusion models reproduce the skewness pattern more closely than the NG-LIM (pattern correlation 0.96 for both vs 0.81; RMSE 0.17 and 0.14 vs 0.26), including the extension of positive skewness from the eastern into the central Pacific and the elongated negative skewness south-east of Indonesia. Their PC1--PC2 scatter also captures the asymmetry along the EP+/CP$-$ axis better than the NG-LIM. Out of sample, on the held-out months, both reach 0.72 (NG-LIM, in-sample: 0.53), equal to the data ceiling, i.e. the correlation between the observed skewness patterns of 1948--2007 and 2008--2022: with a single short record, observations cannot discriminate between these approaches in skill, in line with the sampling-limited regime identified in Figure~2. Pre-training instead halves the spread across training runs and the training time on observations (Tables~S1 and S4).

  \begin{figure}
  \noindent\includegraphics[width=\textwidth]{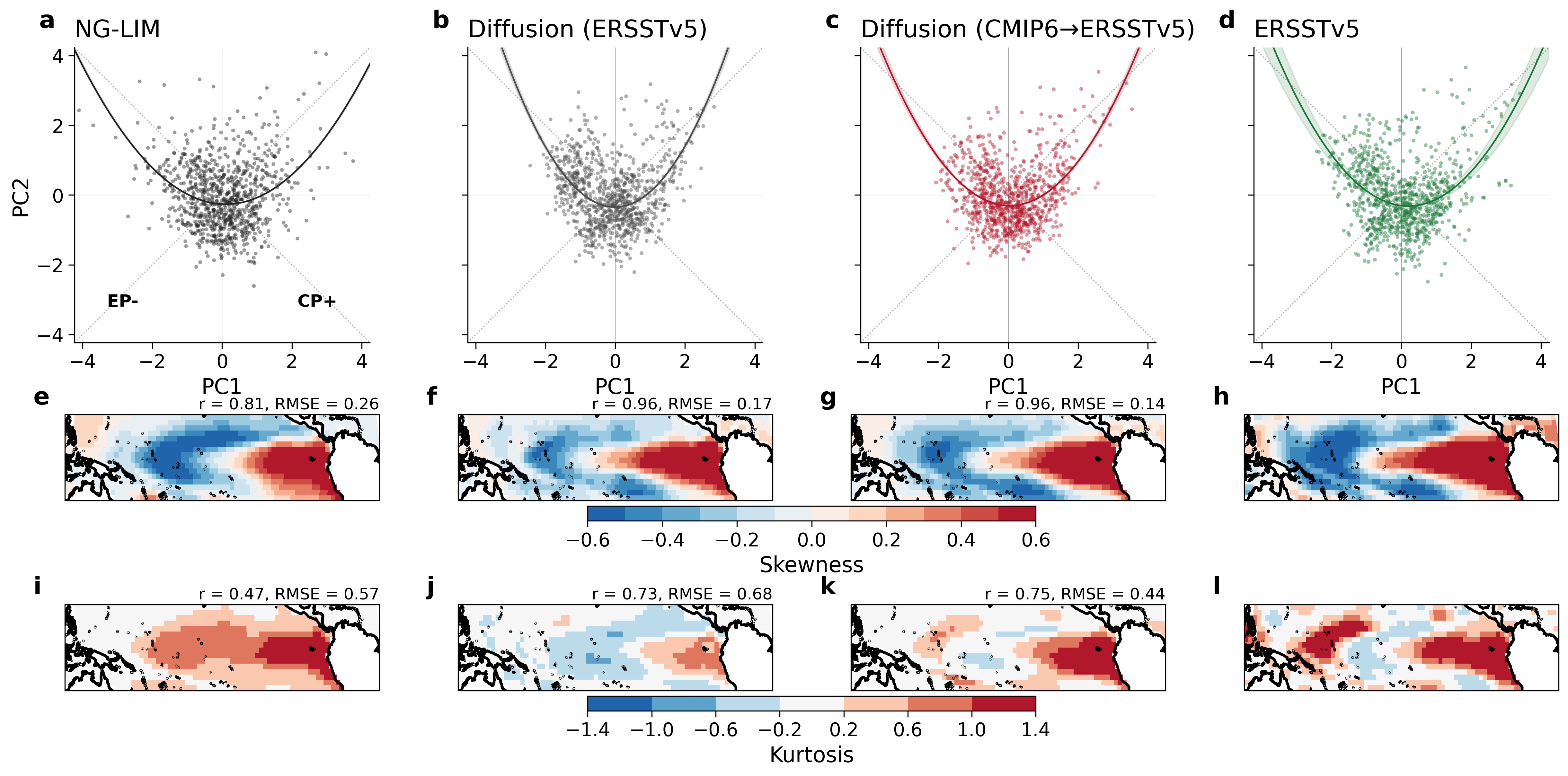}
  \caption{(a--d) Scatter plots of the two leading principal components (PC1 vs PC2) and (e--h) skewness and (i--l) kurtosis maps for (a, e, i) the NG-LIM, (b, f, j) the diffusion model trained directly on the first 720 months of ERSSTv5, (c, g, k) the diffusion model pre-trained on CMIP6 and fine-tuned on the same months, and (d, h, l) the observed record (ERSSTv5, 1948--2022). Positive skewness (red) indicates that warm anomalies reach larger amplitudes than cold anomalies; positive kurtosis indicates heavier tails than a Gaussian. Pattern correlation and RMSE are computed against (h) and (l), the full 1948--2022 record, and are in-sample (see text); Figure~S10 repeats the comparison on the held-out 2008--2022 months.}
  \label{fig:summary}
  \end{figure}

\section{Summary and Discussion}\label{sec:discussion}

Conditional applications of diffusion models, from ensemble forecasting to downscaling, must remain consistent with the unconditional distribution of the training data. Here we tested how faithfully this distribution is learned for tropical Pacific SST in an idealised setting, using synthetic data from a low-order ENSO model with a known generating distribution \cite{martinez-villalobosLowOrderDataDrivenModel2025}. We asked whether a diffusion model recovers its variability patterns, variance partition and non-Gaussian features, how much of these can be learned from a record as short as current reanalyses, and whether regularisation or pre-training can compensate for such limited sampling. The main findings are:

\begin{enumerate}
    \item Given sufficient training data, diffusion models recover these statistics for both Gaussian and non-Gaussian synthetic datasets.
    \item Only the $\mathbf{x}$-prediction parameterisation recovers these statistics robustly as the input dimension grows, consistent with \citeA{liBackBasicsLet2025}.
    \item Reproducing these statistics reliably requires substantially more data than the observational record provides: trained on a record of its length, the model falls short of the scores achieved with longer records. Early stopping, the effective regulariser of diffusion models \cite{farghlyBenignOverfittingDoes2026}, prevents overfitting; the limitation lies instead in the information content of the training record, which leaves the distribution's higher-order moments the least well constrained.
    \item Pre-training on climate simulations improves accuracy mainly for records shorter than ERSSTv5. At ERSSTv5's length, the added value comes mainly from greater stability across random seeds and shorter training runs.
    \item On ERSSTv5, the diffusion model reproduces the observed skewness pattern more closely than the NG-LIM, although this result is subject to the limited length of the observed record \cite{wittenbergAreHistoricalRecords2009}.
\end{enumerate}

Our results show that diffusion models are effective tools for modelling non-Gaussian distributions in climate applications such as ENSO, although not all parameterisations are equally suited \cite{liBackBasicsLet2025}. At observational length, however, the bottleneck is the record rather than the model: neither regularisation nor pre-training can substitute for variability the training record does not sample. Our experiments also highlight two caveats. First, the model interpolates within its training distribution but does not extrapolate beyond it: removing extremes from the training data prevents it from generating them. Second, for very short records, summary statistics can look satisfactory while the model is memorising its training samples, which calls for more sensitive diagnostics.

Several limitations should be acknowledged. Our experiments use a controlled, synthetic setup that excludes non-stationary behaviour such as long-term trends or regime shifts, which is far from the true complexity of the climate system. The amount of data our conclusions require is therefore specific to this framework: it depends on the architecture, the training configuration and the intrinsic dimension of the target. Data requirements are likely larger for higher-dimensional targets, for instance those including subsurface variables. Furthermore, our evaluation framework is purely statistical and unconditional: it assesses whether diffusion models match the low-dimensional manifold of the training distribution. However, it does not evaluate temporal dynamics or the predictive skill that would require a modelling framework that conditions on an initial state. Accordingly, the model does not serve as an ENSO prediction system in its current form (unlike the LIM and NG-LIM). Besides, following \citeA{martinez-villalobosLowOrderDataDrivenModel2025}, the ENSO state is characterised by SSTs alone, a strong simplification of its coupled ocean--atmosphere dynamics \cite{newmanEmpiricalModelTropical2011}.


Overall, our study contributes a controlled framework for verifying what generative models learn from short climate records --- a step that must precede their use for prediction, reconstruction or data assimilation, which all build on the unconditional distribution these models learn. Synthetic data from a known process allow us to (i) establish what a diffusion model has learned against a ground truth, (ii) quantify how much data it requires and separate the model's limitations from those inherent to the record, (iii) detect memorisation that standard summary statistics miss, and (iv) evaluate strategies intended to make training feasible on short records, such as regularisation and pre-training on climate simulations. The methodology is not specific to ENSO and applies to any data-scarce geophysical problem with an inherently chaotic component. Conditional generation is the natural next step. Its skill, too, will be hard to verify on short observational records \cite{verjansLargePotentialPerformancebased2026}, which makes synthetic benchmarks such as the one presented here all the more necessary.

\section*{Open Research Section}

All datasets used in this publication are publicly available via \citeA{martinez-villalobosTropicalPacificSST2026}. Code and instructions to reproduce the results of this manuscript can be found in \url{https://gitlab.earth.bsc.es/es/enso_diff}.

\section*{Conflict of Interest declaration}

The authors declare there are no conflicts of interest for this manuscript.

\acknowledgments

We thank the reviewers for their useful comments and suggestions. V.V. acknowledges funding from the European Union Horizon project EXPECT (grant 101137656).

%
%

\bibliography{DIFF_MANIFOLD.bib}

%
%
%
%
%

\end{document}